# Acoustic Metafluid for Independent manipulation of the Mass Density and Bulk Modulus

Yafeng Bi[1,4], Ping Zhou[1,2,4], Han Jia[2,3], Fan Lu[1], Yuzhen Yang[1], Peng Zhang[1,2] and Jun Yang[*1,2]

[1] Key Laboratory of Noise and Vibration Research, Institute of Acoustics, Chinese Academy of Sciences, Beijing 100190, People's Republic of China

[2] University of Chinese Academy of Sciences, Beijing 100049, People's Republic of China

[3] State Key Laboratory of Acoustics, Institute of Acoustics, Chinese Academy of Sciences, Beijing 100190, People's Republic of China

[4] These authors contributed equally: Yafeng Bi, Ping Zhou

[*] Authors to whom correspondence should be addressed: hjia@mail.ioa.ac.cn;

jyang@mail.ioa.ac.cn;

**Abstract**

Tuning the mass density and bulk modulus independently is the key to manipulate the propagation of sound wave. Acoustic metamaterials provide a feasible method to realize various acoustic parameters. However, the relevant studies are mainly concentrated in air, and the huge impedance difference makes it difficult to directly extend these airborne structures to underwater application. Here, we propose a metafluid to realize independent manipulation of the mass density and bulk modulus underwater. The metafluid is composed of hollow regular polygons immersed in the water. By adjusting the side number of the hollow regular polygons and choosing proper materials, the effective mass density and bulk modulus of the metafluid could be modulated independently. Based on the flexible adjustment method, metafluids with same impedance but different sound velocities are designed and used to realize an underwater impedance-matched gradient index lens. In addition, by combining the proposed metafluid with other artificial structures, acoustic parameters with great anisotropy can be achieved, which is exemplified by the design and demonstration of an impedance-matched underwater acoustic carpet cloak. This work can expand the practicability of underwater metamaterials and pave the way for future potential engineering applications in the practical underwater devices.

# Introduction

Acoustic metamaterials composed of subwavelength components have attracted great interests over the past few decades. Based on different mechanisms, a variety of exotic acoustic parameters have been realized, including the negative mass density and bulk modulus[1,2], zero-refractive-index [3,4], anisotropic mass density[5-8], etc. Combined with transformation acoustics, acoustic metamaterials provide an effective approach to design and implement various devices, such as acoustic cloaks[9-14], acoustic lens[15-17] and acoustic rectifiers[18,19]. However, the required parameters of these devices typically vary continuously with spatial position, which poses a challenge to the parameter modulating ability of acoustic metamaterials. Therefore, acoustic metamaterials with greater ability to independently adjust the mass density and bulk modulus are needed in the realization of acoustic devices.

In recent years, there have been some studies on independent adjustment of mass density and bulk modulus[20-23]. For example, the unit cell made of perforated plate and side pipe is used to control the effective mass density and bulk modulus in air[22]. The variation of the hole in the perforated plate controls the effective mass density while the height of the side pipe modulates the effective bulk modulus. The expansion of dimensions gives the unit cell additional geometric freedom, thus realizing the independent manipulation of mass density and bulk modulus. With the help of the membrane-type acoustic metamaterials, the independent adjustment of mass density and bulk modulus are also achieved by employing the meta-atom with inner sub-cell and outer sub-cell[23]. The inner sub-cell contributes to the bulk modulus and outer sub-cell controls the mass density. However, due to the huge difference of impedance, these unit cells in the air cannot be directly applied to the design of the underwater acoustic devices.

For the underwater acoustic parameter modulation, researchers proposed solid-based pentamode material, which can provide access to tailor its effective density and bulk modulus independently[24-30]. By means of hollowing and lithography technology, two-dimensional (2D) and 3D pentamode materials have been fabricated and demonstrated successively. The thickness of the edge controls the modulus and the mass at the vertex controls the effective mass density. Although the pentamode material can achieve independent control of mass density and bulk modulus, the shear modulus which cannot be eliminated still brings a lot of resonance at low frequency range and affects its broadband performance. The parameters modulation based on pentamode materials is also limited by the complex structure and precise manufacture, which cannot meet the independent, convenient and broadband parameter adjustment in the underwater environment. Thus, achieving an underwater metafluid which can adjust its mass density and bulk modulus simply and independently in a wide frequency range remains a challenge.

In this article, we propose a metafluid consisting of the metal hollow regular polygons immersed in water. By changing the side number of the hollow regular polygons and choosing proper frame materials, the effective mass density and bulk modulus of the metafluid can be adjusted independently. The performance of the metafluid is verified by both simulations and experiments. Then an impedance-matched underwater gradient-index (GRIN) lens is designed to show the ability to independently adjust mass density and bulk modulus. Moreover, an impedance-matched underwater carpet cloak is realized by combining hollow regular polygons and steel strips. The good performance demonstrates that the proposed metafluid has great potential for underwater acoustic applications.

## Theoretical analysis and demonstrations of the unit cell.

The proposed metafluid is made of hollow regular polygons which is arranged in a square lattice and immersed in water. A typical unit cell composed of a hollow regular octagon is presented in Fig.1a to schematically illustrate the design principle. The detailed geometric parameters are marked in the figure, wherein $a$ represents the lattice constant; $r$ is the length from the center to the external vertex of the regular polygon; $h$ is the thickness of the solid shell; $\theta$ is the central angle facing the side and depends on the side number of the regular polygon $N$ ($N=8$, $\theta = 45°$ for the hollow octagon in Fig. 1a).

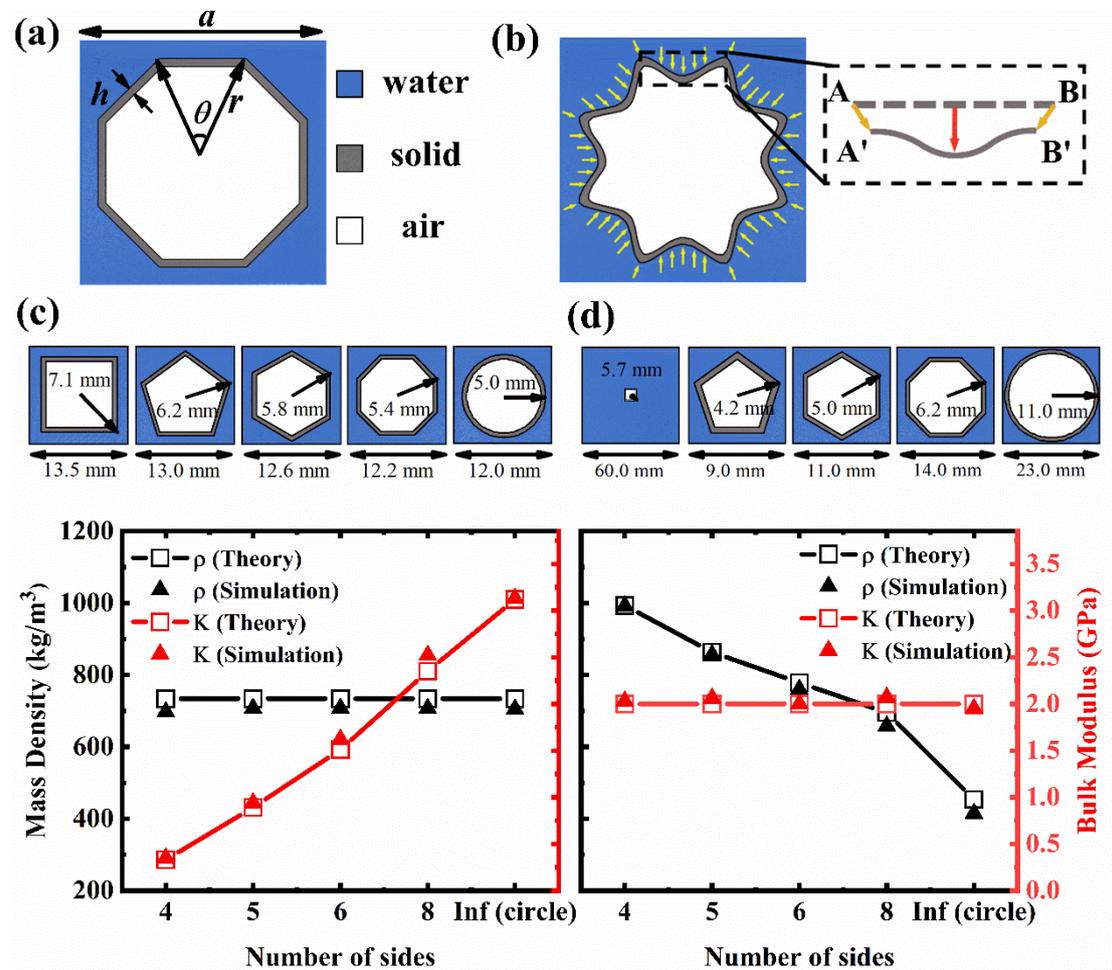

**Fig. 1| Schematic and independent manipulation of the proposed metafluid. a,** Schematic view of the unit cell. A hollow regular polygon (an octagon is chosen as an example) is immersed in water and arranged as a square lattice. **b,** Schematic deformation of the hollow regular polygon under isotropic pressure. Inset: enlarged view of the side deformation in the hollow polygon. **c, d,** Designed metafluids with the same effective mass density (ρ) but different

bulk moduli (K) **(c)** and the same effective bulk modulus but different mass densities **(d)**. The structures and geometric parameters of the metafluids are depicted in the upper panels. All the hollow regular polygons are made of aluminum, and the thickness of all the solid shell is $h=0.5$mm.

In the long wavelength regime, the proposed metafluid can be regarded as a homogeneous medium. The effective mass density of the metafluid can be expressed as:

$$\rho_{eff} = \rho_w \frac{2a^2 - Nr^2 \sin(\theta)}{2a^2} + \rho_s \frac{N\sin(\theta)\left(2\frac{rh}{\sin(\theta)} - \frac{h^2}{\sin^2(\theta)}\right)}{2a^2}, \quad (1)$$

where $\rho_w$ and $\rho_s$ represent the mass densities of water and solid shell, respectively. The air is ignored in equation (1) for the extremely small mass density. On the other hand, the effective bulk modulus is determined by the relative deformation of each component in the unit cell under the external pressure. According to the homogenization theory, the effective bulk modulus of the unit cell could be written as:

$$\frac{1}{B_{eff}} = \frac{1}{a^2}\left(\frac{S_w}{B_w} + \frac{dS_{shell}}{dP}\right), \quad (2)$$

where $S_w$ and $S_{shell}$ are the area of ambient water and hollow regular polygon, respectively; $B_w$ is the bulk modulus of ambient. The first item on the right side of equation (2) represents the relative deformation of ambient water while $dS_{shell}/dP$ is the relative decrease of the area from the hollow regular polygon. When the external static pressure is applied, the polygon shell will deform uniformly, as is illustrated in Fig. 1b. The deformation consists of two parts: translational deformation originating from the displacement of the vertex (orange arrows, inset of Fig. 1b) and flexural deformation from the bend of the side (red arrow, inset of Fig. 1b). Then the relative area variation of the hollow regular polygon can be obtained by accumulating these two parts and expressed as:

$$\frac{dS_{shell}}{dP} = N\left(\frac{L^5}{60Eh^3} + \frac{L^3h^2(1+\mu)}{4Eh^3}\right) + NL\frac{(1+\mu)r}{E}\frac{\left(\frac{h}{r}-2\right)\frac{h}{r}+2(1-\mu)}{\left(\frac{h}{r_1}-2\right)\frac{h}{r}}\cos\left(\frac{\theta}{2}\right), \quad (3)$$

where $E$ and $\mu$ are the Young's modulus and Poisson's ratio of the solid shell, respectively; and $L = 2\left(r - \frac{h}{\sin(\theta)}\right)\sin\left(\frac{\theta}{2}\right)$ is the inner side length of the solid hollow regular polygon. The detailed derivation is presented in Supplementary Note 1. The effective bulk modulus of the unit cell can be obtained by substituting equation (3) into equation (2). From above derivations and equations, it can be seen that the effective mass density and bulk modulus can be flexibly modulated by selecting proper materials and adjusting the geometry parameters of the hollow regular polygon. This cooperated adjustment mode provides an avenue to independently modulate the effective mass density and bulk modulus of the metafluid.

**Simulations and verifications for the impedance-matched metafluid.**

Based on the above theoretical model, we have carried out some examples to demonstrate the ability to independently regulate the effective parameters of the proposed metafluid. The material of the hollow regular polygon is chosen as aluminum. Firstly, we designed five unit cells with the same mass density but different bulk moduli. The detailed shape and geometric parameters are shown in the top panel of Fig.1c. The volume fractions of each component are the same for these units, but the side number of the regular polygon is changed. The mass densities and bulk moduli of the designed five units can be calculated using the equations (1) and (2). The results are shown as hollow squares in Fig. 1c. It can be observed that these five unit cells have the same density ($\rho$=734 kg/m$^3$) but the corresponding bulk moduli increase with the number of sides. Then, we designed another five unit cells with the same bulk modulus but different mass densities. The detailed shape and geometric parameters are shown in the top panel of Fig.1d and the corresponding effective parameters are plotted as hollow squares. As can be seen from the results, these five unit cells have the same bulk modulus (K=2 GPa). With the increase of the side number, the corresponding volume fraction of the solid shell decreases so that the mass density decreases gradually. To verify the accuracy of theoretical calculation, we further retrieved the effective mass densities and bulk moduli of these unit cells by simulations. The simulated results are shown as solid symbols in Fig. 1c,d for comparison. It can be seen that the simulated results match well with the analytical results, which proved the validity of the proposed theoretical model. Above results verified the great parameter modulation ability resulting from the multi-degree of freedom adjustment in the proposed metafluid.

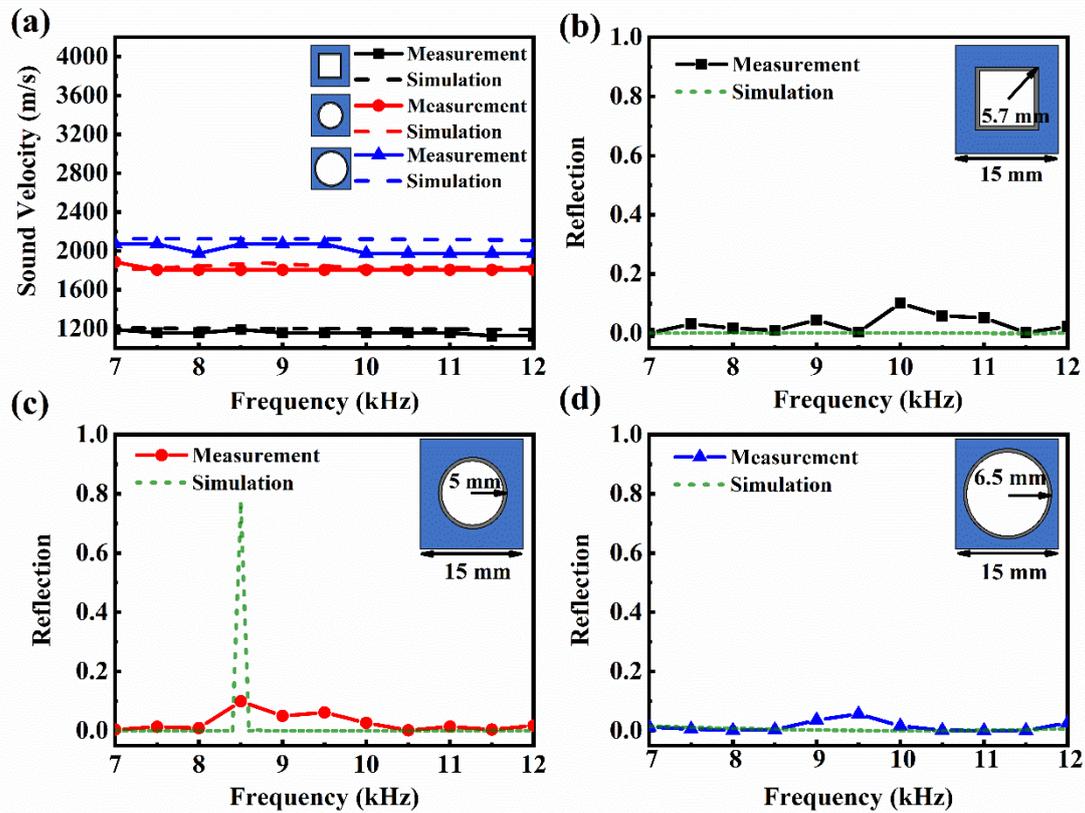

**Fig. 2| Design and demonstration of the impedance-matched metafluids with different sound velocities. a**, Sound velocities of three designed metafluids from 7 kHz to 12 kHz. **b-d**, The energy reflections of the designed metafluid layers composed of the hollow square unit cell **(b)**, the small hollow circular unit cell **(c)** and large hollow circular unit cell **(d)**, respectively. The hollow square frame is made of steel and the hollow circular frame is made of aluminum. The thickness of all the solid shell is $h$=0.5 mm.

During the sound transmission between different media, sound velocity

$c = \sqrt{B/\rho}$ and characteristic impedance $Z = \sqrt{\rho B}$ of the media determine the propagation direction and energy transmission efficiency, respectively. Thus, the materials with the same impedance and different sound velocities are always desired in the design of acoustic devices. If the mass density and the bulk modulus can be adjusted independently, the impedance-matched acoustic materials with different sound velocities can also be realized. Here, we utilize the proposed model to design three metafluids which are impedance matched with water but have different sound velocities. The hollow regular polygon of three metafluids are composed of steel hollow squares (N=4) and aluminum hollow circles (N=∞). The detailed geometric configuration and parameters are shown in the insets of Fig. 2b,c,d. We fabricated three metafluid layers using these three unit cells, and measured the energy reflections and sound velocities in an anechoic tank. The details on the sample fabrication and experimental setup can be found in Methods. As can be seen from Fig. 2b,c,d, the reflection coefficients of three metafluid layers are all below 0.15 in the measured frequency range, which means that the impedances match well with that of water. On the other hand, the measured sound velocities are plotted in Fig. 2a. It can be observed that the sound velocities of three metafluids are about 1200m/s, 1800m/s and 2100m/s, respectively. The sound velocities of all three metafluids keep stable in the measured frequency range. We also calculated the sound velocities and energy reflections by simulations. The corresponding results are shown in Fig. 2a,b,c,d with dashed lines for comparison. It can be noted that the simulated results match well with the measured results expect for the narrow band resonance in Fig. 2c, which is hard to detect in experiments.

**Design and implementation of the impedance-matched GRIN lens.**

With the ability to independently adjust the effective acoustic parameters, the proposed metafluid is promising for applications in underwater impedance-matched acoustic devices. We use the proposed metafluid to design an impedance-matched underwater GRIN lens, which is 0.15 m thick along the propagation direction (*x* axis) and 0.78 m wide in the vertical direction ($y$ axis). A hyperbolic secant sound velocity distribution along the $y$ axis is chosen and shown as the black dashed line in Fig. 3b. Only the negative half part along the $y$ axis is presented as the sound velocity distribution is symmetric about the center of the lens. To realize these sound velocities, we divide the whole lens into 10×52 squares with the side length of 15 mm. The continuous sound velocity is divided into six parts whose spatial distributions along the $y$ axis are marked with black horizontal arrows in Fig. 3b. It is worth noticing that the fourth part has the same sound velocity as the water, therefore, we directly fill this part with water. Then we designed five unit cells, including three hollow aluminum circulars and two steel squares, to realize another five discrete sound velocities. The lattices of the unit cells are the same as these segmented squares. The detailed geometric parameters of the unit cells are presented in Supplementary Table S1. The calculated sound velocities of these unit cells are labelled with corresponding symbols in Fig. 3b. As can be seen from the results, the sound velocities of these unit cells match well with the required sound velocities. Moreover, the transmission coefficients of these unit cells in the wide frequency range are all above 0.98 (see Supplementary Note 2), which promises the impedance-matched property of the device.

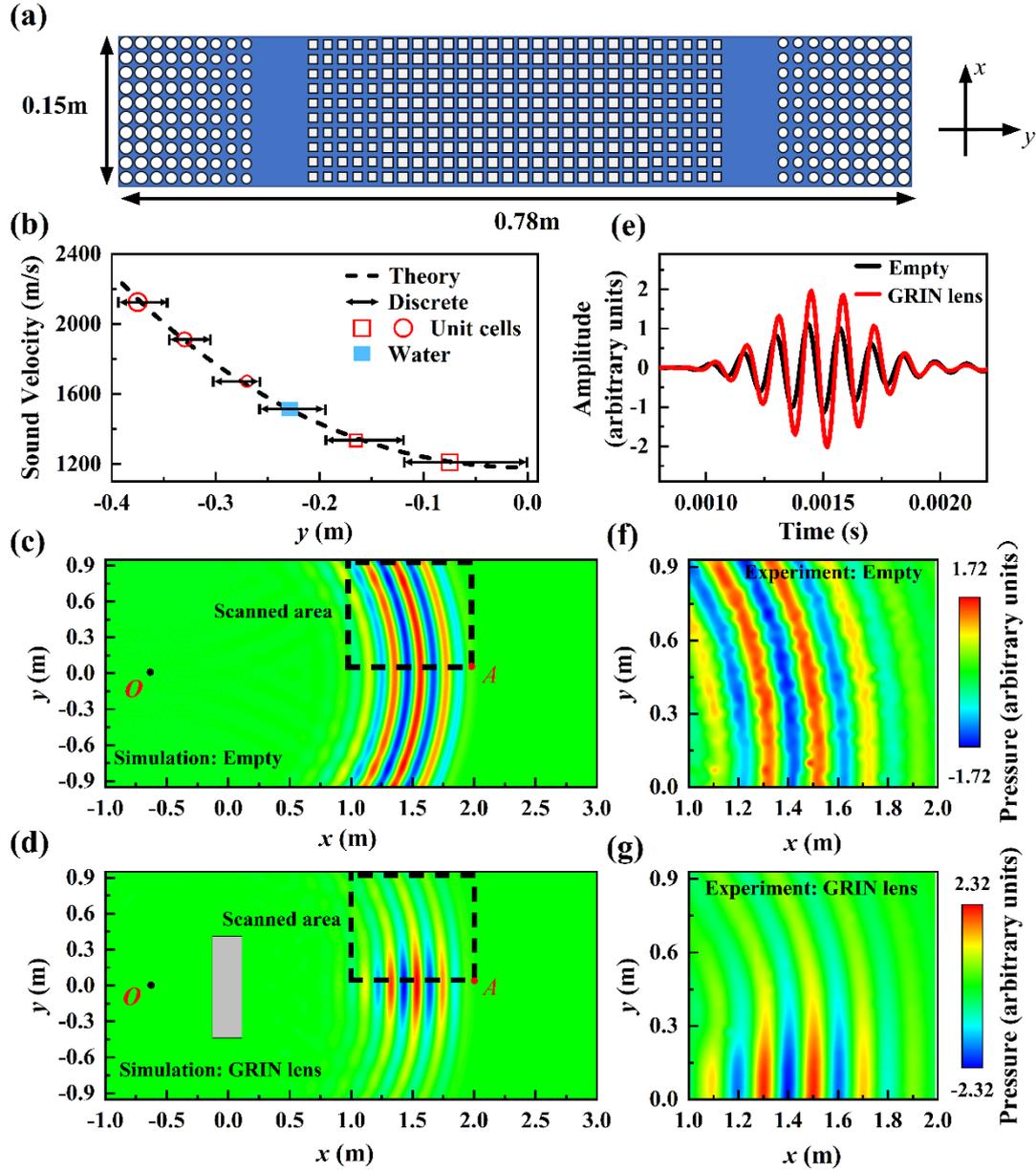

**Fig. 3| Design and demonstration of the GRIN lens. a,** Schematic view of the GRIN lens. **b,** Desired, discrete and realized sound velocity distribution along the *y*-axis. **c, d,** Simulated acoustic fields of the empty anechoic tank (c) and GRIN lens (d) at 7 kHz. The scanned area is marked with black dashed box, wherein point A represents the particular point to extract the time domain signals. **e,** Extracted signals from point A marked in (c) and (d). **f, g,** Experimentally measured acoustic fields of the corresponding empty anechoic tank (f) and GRIN lens (g) at 7 kHz.

The GRIN lens profile filled with the designed metafluids is shown in Fig. 3a. Both simulations and experiments are performed to demonstrate the validity of the designed GRIN lens. A point source is fixed on the focal point *O* (58 cm from the center of the GRIN lens). A Gaussian-modulated pulse centered at 7 kHz is emitted from the source. The simulated pressure field distributions after 1.05 ms are extracted and shown in Fig. 3c,d. As can be seen in Fig. 3c, the wave spreads evenly without GRIN lens. After travelling through the GRIN lens in Fig. 3d, the originally circular wavefront is transformed into a plane shape propagating along the *x*-axis. We further experimentally measured the acoustic pressure fields. A rectangle area marked with black dash lines in Fig. 3c,d is chosen as the scanned area, which is 0.99 m wide and 0.93 m high. The measured instantaneous pressure fields are shown in Fig. 3f,g. In the empty anechoic water tank, the cylindrical wave spreads evenly when it comes into the measured area. After passing through the GRIN lens, the transmitted wave becomes a plane shape along the *x*-axis. Moreover, we extracted the measured time-domain signals at point A (marked with red dots in Fig. 3c,d), which are presented in Fig. 3e. It is obviously that the amplitude becomes much larger after passing through the GRIN lens, which is due to the energy concentration resulting from the contraction of the wave front. What is more, the pulse keeps the time domain Gaussian shape, which confirms the broadband property of the GRIN lens. The measured acoustic pressure fields at 12 kHz are also shown in Supplementary Fig. S4.

**Cooperative design and demonstration of the acoustic carpet cloak.**

In addition, the proposed hollow regular polygons can also be combined with other artificial structures to achieve more exotic underwater acoustic parameters. Here, we use the hollow regular polygons and the steel strips to form an impedance-matched anisotropic metafluid and design an underwater acoustic carpet cloak. The geometric configuration of the carpet cloak is schematically shown in Fig. 4a. The blue region is the background water while the yellow region represents the target to be hidden. The carpet cloak covers the target. By using the transformation acoustics, the mass densities and bulk modulus along the principal axes and required by the cloak are calculated and plotted as dashed lines in Fig. 4b,c, respectively. It can be observed that the required mass densities along the principal axes show significant anisotropy while the required bulk modulus is isotropic. The calculation details are shown in Supplementary Note 3.

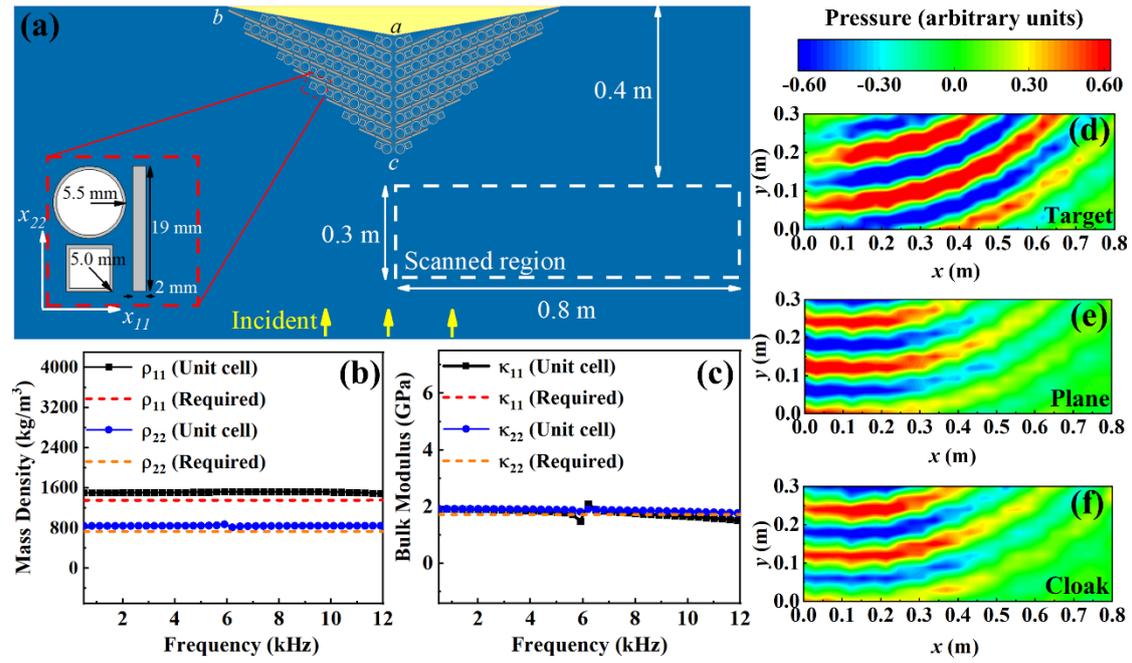

**Fig. 4| Design and demonstration of the impedance-matched underwater carpet cloak. a,** Schematics of the experimental setup. The carpet cloak sample with a size of *a*=35 mm, *b*=198.3 mm and *c*=166.7 mm. The red dotted box displays the unit cell designed to implement the carpet cloak. The principal axes of the unit cell are marked with $x_{11}$ and $x_{22}$. The relative position of sample and scanned area is marked on the right of the panel. The yellow arrows indicate the direction of the incident wave. **b, c,** The required and simulated mass density (b) and bulk modulus (c) along the principal axes of the designed unit cell. **d-f,** The measured reflected pressure fields of the soft target (d), reflecting plane (e) and carpet cloak (f) at 12 kHz.

To realize the required acoustic parameters, a unit cell composed of a hollow aluminum circular, a hollow aluminum square and a steel strip is designed and shown in the red dashed frame in Fig. 4a. The size of the whole lattice is 15 mm by 20 mm. In the left side of the unit cell, two aluminum hollow structures are arranged vertically: one is a hollow circular (*r*=5.5 mm, *h*=0.5 mm) and the other is a hollow square (*r*=5.0 mm, *h*=0.5 mm). The two hollow structures form the low-density part and provide a suitable bulk modulus in the unit cell. The steel strip is placed in the right side of the unit cell and forms the high-density part of the unit cell. Combining these two parts, we obtained the anisotropic metafluid required by the carpet cloak. The effective mass densities and bulk moduli along two principal axes of the designed unit cell are retrieved by simulations and plotted as symbols in Fig. 4b,c, respectively. It can be observed that the effective parameters of the unit cell match well with the parameters required by the carpet cloak in a wide frequency range. A small fluctuation near 6 kHz is caused by the resonance of the circular frame.

To demonstrate the effectiveness of the designed carpet cloak, we carried out experiments in the anechoic water tank. The omnidirectional incident wave spreading from the bottom to top hits the samples and reflects to the opposite side. The scanning region is marked with white dashed line in Fig. 4a. See Methods for more details about the measurement. The measured instantaneous reflected pressure fields of the target, plane and cloak at 12 kHz are shown in Fig. 4d-f. The scattered wave from the target is shown in Fig. 4d. Due to the slope of the triangular target, it is clear that the wave propagates obliquely to the lower right, so that the energy below the target is significantly weaker than the energy in the oblique region. In contrast, the scattered wave from the reflective plane focuses on the backscattering direction as shown in Fig. 4e. After covering the target with carpet cloak, the scattered wave returns to the backscattering direction again, as shown in Fig. 4f. Comparing these three panels, the scattered wave field distribution of the carpet cloak is much closer to that of the plane, which indicates that the acoustic signature of the target is almost cancelled. The cloaked target mimics the reflective plane successfully. The corresponding time domain simulated pressure fields are presented in Supplementary Note 3, which maintains the same sound field characteristics as the experimental results.

## Conclusion and outlook

In conclusion, a broadband underwater metafluid was proposed to independently adjust the mass density and bulk modulus. The metafluid is comprised of hollow regular polygons immersed in the water. Both simulated and experimental results confirm the effective parameter modulating approach and show the adjustment ability of multi-degrees of freedom in designing underwater metafluid. Utilizing the proposed method, we have designed an impedance-matched underwater GRIN lens. The measured sound pressure distributions show that the circular wavefront is smoothly transformed into a plane wavefront after passing through the GRIN lens. Moreover, the proposed metafluid can also be combined with existing artificial structures to achieve cooperative manipulation. By combining hollow aluminum polygons and steel strips, we designed an anisotropic metafluid and used it to realize an impedance-matched underwater carpet cloak. The experimental results confirm that the carpet cloak can conceal the target in a wide frequency range.

The design method reported in this work opens new avenues for both design and applications of future acoustic metamaterials. The collaborative adjustment mode based on geometry and material parameters not only endows great parameter adjustment ability, but also immensely expands the achievable acoustic parameter range. It is expected to provide solutions for applications such as nondestructive testing, biomedical ultrasonography, and acoustic communication based on impedance matching. Additionally, the proposed metafluid is low cost and easy to design. In general, this work expands the practicability of underwater metamaterials and brings great potential engineering applications in the practical underwater devices. In the future work, the combination of richer material categories and geometric shapes may be further explored for more acoustic applications.

## Methods

**Numerical simulations.**

The simulations are conducted by the acoustic-solid interaction module in a finite element package (COMSOL Multiphysics). In the simulations, the material parameters of water are set as: $\rho=1000$ kg/m$^3$, K=2.19 GPa; the material parameters of air are set as: $\rho=1.21$ kg/m$^3$, $c=343$ m/s; the material parameters of aluminum are set as: $\rho=2700$ kg/m$^3$, E=70 GPa, $\mu=0.33$; the material parameters of steel are set as: mass $\rho=7850$ kg/m$^3$, E= 205 GPa, $\mu=0.28$. When calculating the sound velocity and energy reflection, the frequency domain simulation is conducted. The unit cell is placed in the waveguide filled with water. The inlet and outlet of the waveguide are set as plane wave radiation. When calculating the pressure filed, the time domain simulation is conducted. The background fluid is water, and the boundaries of the simulated area are set as plane wave radiation.

**Sample fabrications.**

All the designed unit cells are fabricated into hollow pipes with a length of 1000 mm. Both ends of these small pipes are sealed with glass cement to prevent water from penetrating. Then, all these pipes are inserted in two identical perforated plate supports for fixing. Finally, the whole system is submerged in water to form a quasi-2D sample. Three impedance matching layer samples are made, and each sample contains 7×35 unit cells. The GRIN lens sample is shown in Supplementary Fig. S3, and the carpet cloak sample is shown in Supplementary Fig. S7.

**Experimental measurements.**

In the measurement of sound velocities, the impedance-matched layer sample was submerged 3m below the surface in an anechoic tank. An omnidirectional cylindrical transducer was placed in front of the samples (about 3.5 m away from the sample). Two hydrophones (Type 8103, B&K) were fixed 0.2 m away from the surfaces of the sample to measure the times-domain signals: one was in front of the sample while the other was in back of the sample. A series of short Gaussian pulses (7 kHz to 12 kHz with 0.5 kHz step) were emitted to test the time delay between these two hydrophones. All the acoustic signals were emitted and received by a signal acquisition (PXI-6115, NI) to estimate the delay more accurately. Then the sound velocity in each sample could be simply calculated by the distance and delay time. The energy reflection of each sample was obtained by the measuring the acoustic wave with the fixed hydrophone in front of the sample when the samples were loaded and unloaded. Then the ratio of the reflected energy to the incident energy could be calculated.

In the scanning of pressure fields, the samples are immersed in water and an omnidirectional cylindrical transducer is placed at the designated position. The sound source generates short Gaussian pulse (center frequency 7 kHz, bandwidth 1 kHz in the GRIN lens experiment; center frequency 12 kHz, bandwidth 2 kHz in the carpet cloak experiment). The propagation of the wave is measured in time-domain using two hydrophones (Type 8103, B&K): one is fixed near the transducer as the monitoring hydrophone while the other scans the area step by step (3 cm per step). All the emitting and receiving acoustic signals are analyzed by a multi-analyzer system (Type 3560,

B&K).

# References

[1] Lee, S. H. et al. Acoustic metamaterial with negative density. *Phys. Lett. A*. **373**, 4464-4469 (2009).

[2] Lee, S. H. et al. Acoustic metamaterial with negative modulus. *J. Phys-Condens. Mat.* **21**, 175704 (2009).

[3] Huang, X. et al. Dirac cones induced by accidental degeneracy in photonic crystals and zero-refractive-index materials. *Nat. Mater.* **10**, 582–586 (2011).

[4] Dubois, M. et al. Observation of acoustic Dirac-like cone and double zero refractive index. *Nat. Commun*. **8**, 14871 (2017).

[5] Torrent, D. & Sanchez-Dehesa, J. Acoustic cloaking in two dimensions: a feasible approach. *N. J. Phys*. **10**, 063015 (2008).

[6] Torrent, D. & Sánchez-Dehesa, J. Anisotropic mass density by two-dimensional acoustic metamaterials. *N. J. Phys*. **10**, 023004 (2008).

[7] Popa, B. I., Wang, W., Konneker, A. & Cummer, S. A. Anisotropic acoustic metafluid for underwater operation. *J. Appl. Phys*. **109**, 054906 (2011).

[8] Popa, B. I. & Cummer, S. A. Design and characterization of broadband acoustic composite metamaterials. *Phys. Rev. B*. **80**, 174303 (2009).

[9] Hu, W. et al. An experimental acoustic cloak for generating virtual images. *J. Acoust. Soc. Am*. **139,** 024911 (2016).

[10] Zigoneanu, L., Popa, B. I. & Cummer S. A. Three-dimensional broadband omnidirectional acoustic ground cloak. *Nat. Mater.* **13**, 352 (2014).

[11] Bi, Y. et al. Design and demonstration of an underwater acoustic carpet cloak. *Sci. Rep.* **7**, 705 (2017).

[12] Chen, Y. et al. Broadband solid cloak for underwater acoustics, *Phys. Rev. B*. **95**, 180104 (2017).

[13] Kerrian, P. A. et al. Development of a perforated plate underwater acoustic ground cloak. *J. Acoust. Soc. Am.* **146**, 2303 (2019).

[14] Zhou, P. et al. Underwater carpet cloak for broadband and wide-angle acoustic camouflage based on three-component metafluid. *Phys. Rev. Appl*. **18**, 014050 (2022).

[15] Zigoneanu, L., Popa, B. I., Starr, A. F. & Cummer, S. A. Design and measurements of a broadband two-dimensional acoustic metamaterial with anisotropic effective mass density. *J. Appl Phys*. **109**, 054906 (2011).

[16] Martin, T. P. et al. Transparent gradient-index lens for underwater sound based on phase advance. *Phys. Rev. Appl*. **4**, 034003 (2015).

[17] Li, J. et al. Experimental demonstration of an acoustic magnifying hyperlens. *Nat. Commun.* **8**, 931–934 (2009).

[18] Liang, B. et al. An acoustic rectifier. *Nat. Mater.* **9**, 989–992 (2010).

[19] Boechler, N., Theocharis, G. & Darai, C. Bifurcation-based acoustic switching and rectification. *Nat.Mater.* **10**, 665–668 (2011).

[20] Ding, Y., Liu, Z., Qiu, C. & Shi, J. Metamaterial with simultaneously negative bulk modulus and mass density. *Phys. Rev. Lett*. **99**, 093904 (2007).

[21] Oh, J. H., Kwon, Y. E., Lee, H. J. & Kim, Y. Y. Elastic metamaterials for independent realization of negativity in density and stiffness. *Sci. Rep.* **6**, 23630 (2016).

[22] Yang, Y. et al. Impedance-matching acoustic bend composed of perforated plates and side pipes. *J. Appl Phys.* **122**, 054502 (2017).

[23] Koo, S., Cho, C., Jeong, J. & Park, N. Acoustic omni meta-atom for decoupled access to all octants of a wave parameter space. *Nat. Commun.* **7**, 13012 (2016).

[24] Norris, A. N. Acoustic metafluids. *J. Acoust. Soc. Am.* **125**, 839-849 (2009).

[25] Norris, A. N. & Nagy, A. J. Acoustic metafluids made from three acoustic fluids. *J. Acoust. Soc. Am.* **128**,1606-1616 (2010).

[26] Hladky-Hennion, A. C. et al. Negative refraction of acoustic waves using a foam-like metallic structure. *Appl. Phys. Lett.* **102**, 144103 (2013).

[27] Kadic, M. et al. On the practicability of pentamode mechanical metamaterials. *Appl. Phys. Lett.* **100**, 191901 (2012).

[28] Tian, Y. et al. Broadband manipulation of acoustic wavefronts by pentamode

metasurface. *Appl. Phys. Lett.* **107**, 221906 (2015).

[29] Cai, X. et al. The mechanical and acoustic properties of two-dimensional pentamode metamaterials with different structural parameter. *Appl. Phys. Lett.* **109**, 131904 (2016).

[30] Sun, Z. et al. Quasi-isotropic underwater acoustic carpet cloak based on latticed pentamode metafluid. *Appl. Phys. Lett*. **114**, 094101 (2019).

## Acknowledgements

This work is supported by the Key-Area Research and Development Program of Guangdong Province (Grant No. 2020B010190002), the National Natural Science Foundation of China (Grant No. 11874383, 12104480), the IACAS Frontier Exploration Project (Grant No. QYTS202110).

## Author contributions

Y.B., H.J. and P.Z. conceived the idea. Y.B. performed the numerical simulations and theoretical analyses. Y.B., P.Z. and Y.Y. carried out all measurements. J.Y. supervised the whole project. P.Z., H.J. and Y.B. wrote the manuscript and the Supplementary Information. All authors contributed to discussions of the results and the manuscript.

## Competing interests

The authors declare no competing interests.

## Supplementary Information

Supplementary Information is available.